\documentclass[10pt]{article}
\usepackage{graphicx}
\usepackage{hyperref}
\usepackage{amssymb}
\usepackage{subcaption}
\usepackage{cancel}

\newcommand{\jpsi}{\rm J/$\psi$}

\newcommand{\ccbar}{\ensuremath{\mathrm{c\overline{c}}}}

\newcommand{\Dzero}{\ensuremath{\mathrm{D^{0}}}}

\newcommand{\pp}{pp}
\newcommand{\pPb}{p--Pb}
\newcommand{\PbPb}{Pb--Pb}

\newcommand{\GeVc}{GeV/$c$}

\newcommand{\snn}{\ensuremath{\sqrt{s_{\rm NN}}}}

\newcommand{\pt}{\ensuremath{p_{\rm T}}}

\newcommand{\antikt}{anti-\ensuremath{k_{\mathrm{T}}}}
\newcommand{\kt}{\ensuremath{k_{\mathrm{T}}}}

\def\Title#1{\begin{center} {\Large #1 } \end{center}}
\def\Author#1{\begin{center}{ \sc #1} \end{center}}
\def\Address#1{\begin{center}{ \it #1} \end{center}}

\newcommand\pubblock{\rightline{\begin{tabular}{l} Proceedings of the Fifth Annual LHCP\\ \pubnumber\\
         \pubdate  \end{tabular}}}

\newenvironment{Abstract}{\begin{quotation} \begin{center} 
             \large ABSTRACT \end{center}\bigskip 
      \begin{center}\begin{large}}{\end{large}\end{center} \end{quotation}}

\newenvironment{Presented}{\begin{quotation} \begin{center} 
             PRESENTED AT\end{center}\bigskip 
      \begin{center}\begin{large}}{\end{large}\end{center} \end{quotation}}

\def\beq{\begin{equation}}
\def\eeq#1{\label{#1}\end{equation}}
\def\eeqn{\end{equation}}

\def\beqa{\begin{eqnarray}}
\def\eeqa#1{\label{#1}\end{eqnarray}}
\def\eeqan{\end{eqnarray}}

\let\bar=\overbar

\def\Dslash{\not{\hbox{\kern-4pt $D$}}}
\def\dslash{\not{\hbox{\kern-2pt $\del$}}}

\def\msb{{\bar{\ssstyle M \kern -1pt S}}}

\usepackage{color}

\newcommand\pubnumber{ }

\newcommand\pubdate{\today}

\def\affiliation{
On behalf of the ALICE Collaboration, \\
Physics Department, Yale University, New Haven, CT 06511, USA}

\def\support{\footnote{Work supported by the U.S. Department of Energy under grant number DE-SC0004168.}}

\begin{document}

\large
\begin{titlepage}
\pubblock

\vfill
\Title{New results on jets and heavy flavor in heavy-ion collisions with ALICE}
\vfill

\Author{Salvatore Aiola\support}
\Address{\affiliation}
\vfill
\begin{Abstract}
The large statistics accumulated during the LHC Run-1 and Run-2 have provided the unique opportunity to study the properties of the Quark-Gluon Plasma
using rare hard probes. The ALICE Collaboration has presented new results on the jet shape observables in \pPb\ and \PbPb\ collision. 
Using data from Run-2, the elliptic flow of open and hidden charm mesons have been measured with much better precision,
confirming that charm quarks participate in the collective dynamic expansion with the medium. In \pPb\ collisions the absence of jet quenching 
correlated with high event activity has been confirmed by a self-normalized semi-inclusive hadron+jet measurement.
\end{Abstract}
\vfill

\begin{Presented}
The Fifth Annual Conference\\
 on Large Hadron Collider Physics \\
Shanghai Jiao Tong University, Shanghai, China\\ 
May 15-20, 2017
\end{Presented}
\vfill
\end{titlepage}
\def\thefootnote{\fnsymbol{footnote}}
\setcounter{footnote}{0}
%

\normalsize 


\section{Introduction}
A comprehensive and far-reaching heavy-ion programme is part of the scientific goals of the CERN Large Hadron Collider.
The large statistics accumulated during the LHC Run-1 and Run-2 have given access to an unprecedented number of observables connected to rare high-$Q^2$ processes.
Such hard probes are rare enough to be singled out from the large bulk of particles produced in ultra-relativistic heavy-ion collisions. By interacting with the Quark-Gluon Plasma (QGP) these self-produced probes
become carriers of crucial pieces of information about the transport properties of the QGP.

The suppression of the jet production cross section in ultra-relativistic heavy-ion collisions has been established both at RHIC~\cite{STAR:2017a} and at the LHC~\cite{ALICE:2015a}. The suppression is interpreted as the result of the interaction of the hard-scattered partons 
with the QGP, which dampens the momentum of the hard parton via radiative and collisional energy loss (\emph{jet quenching}). A number of theoretical models, such as JEWEL~\cite{Zapp:2014} and YaJEM~\cite{Renk:2013}, can reproduce
the main features observed in data~\cite{ALICE:2015a}; however the exact determination of the parameters of the energy loss as well as the details of the energy-loss mechanism is still pending~\cite{Burke:2013}. Some additional constraints come from the observation (or lack thereof) of
the modification of the internal structure of the jets. A number of observables has been proposed. The ALICE Collaboration is active in this area and some of the latest results have been presented at this conference.

Heavy-flavor (charm and beauty) particles have been used as probes of the QGP both at RHIC and at the LHC. What makes heavy-flavor partons special is their large mass which plays the role of a hard scale, 
independent of the momentum.
As a consequence, perturbative Quantum Chromo-Dynamics (pQCD) methods are effective in calculating the production cross-section in vacuum, i.e. in \pp\ collisions, down to zero momentum, notwithstanding large uncertainties~\cite{ALICE:2017a}.
Furthermore, the non-zero mass of charm and beauty has an impact in their interaction with the QGP, which has been investigated theoretically~\cite{Dokshitzer:2001}.
Therefore, studying the dynamics of heavy quarks in the QGP can put important additional constraints on its transport properties.
At this conference the ALICE Collaboration has presented an updated measurement of the azimuthal anisotropy of D-meson and \jpsi\ production, using the larger statistical samples available from Run-2.

In recent years, collective effects usually associated with the formation of the QGP have been observed in a subset of \pPb~\cite{CMS:2013b} and in \pp~\cite{ALICE:2017b} collisions, characterised by a very large event multiplicity or event activity (EA). These observations could be taken as indication that the smallest drop
of QGP is formed in large EA collisions, however no unambiguous proof has been shown yet. In particular, no unexpected effects have been observed using hard probes (jets or heavy-flavor particles). In these proceedings we present two recent measurements from ALICE.

The ALICE detector and its performance are described elsewhere~\cite{ALICE:2008, ALICE:2014b}. We refer the interested reader to those publications.

\section{\PbPb\ collisions}

\subsection{Nsubjettiness}
\label{sect:nsubjettiness}
If two partons within the same parton shower are separated by a distance smaller than the tranvserse resolution scale of the medium, they interact coherently with the QGP as a color singlet (color coherence). 
Conversely, if they are separated by larger distances they interact incoherently with the medium as independent color charges.
The transverse resolution scale can be characterized with a critical angle $\theta_{\rm c}$. This angle can be related to the medium decoherence parameter $\Delta_{\rm med}$, which enters the calculation of the amplitude of the antenna emission, via Eq.~\ref{eq:delta_med}~\cite{Mehtar-Tani:2012, Casalderrey-Solana:2013}.
\begin{equation}
\Delta_{\rm med} \simeq 1 - e^{-\frac{1}{12}\hat{q}Lr^2_{\bot}} \equiv 1 - e^{-(\theta/\theta_{\rm c})^2}
\label{eq:delta_med}
\end{equation}
Here $\hat{q}$ is the medium transport parameter, $L$ is the medium's
length and $r_{\bot}$ is the jet's transverse extension in the medium. $\theta$ is the jet's opening angle, which is given by
the largest antenna (the two substructures with the largest angular separation) in the jet which in a vacuum
corresponds to the first splitting. Recent theoretical work has highlighted the sensitivity of two-pronged jets
to coherence effects in the QGP~\cite{Mehtar-Tani:2016}.

The nsubjettiness~\cite{Thaler:2010} $\tau_{\rm N}$ is a measure of how $N$-cored a jet's substructure is. It is defined in Eq.~\ref{eq:nsubjettiness}, where
$i$ denotes each jet constituents, $N$ is the number of cores found in the jet, $M$ is the number of jet constituents, $R$ is the jet resolution parameter used in the \kt\ jet finding algorithm
and $\Delta R_{j,i}$ is the distance in the $(\eta,\phi)$ plane between the track $i$ and the jet axis core $j$.
\begin{equation}
\tau_N = \frac{\sum^M_{i=1}p_{{\rm T},i}\min(\Delta R_{1,i},\Delta R_{2,i},...,\Delta R_{N,i})}{\sum^M_{i=1}p_{{\rm T},i}R}
\label{eq:nsubjettiness}
\end{equation}
A jet with $\tau_N$ approaching $0$ 
has $N$ or fewer definite cores; conversely a $\tau_N$ value
approaching unity indicates the presence of at least $N + 1$ substructures. 
It follows that the ratio $\tau_2/\tau_1$ is sensitive to jets that have exactly two definite hard cores. 

\begin{figure}[tb]
\centering
\includegraphics[width=.5\textwidth]{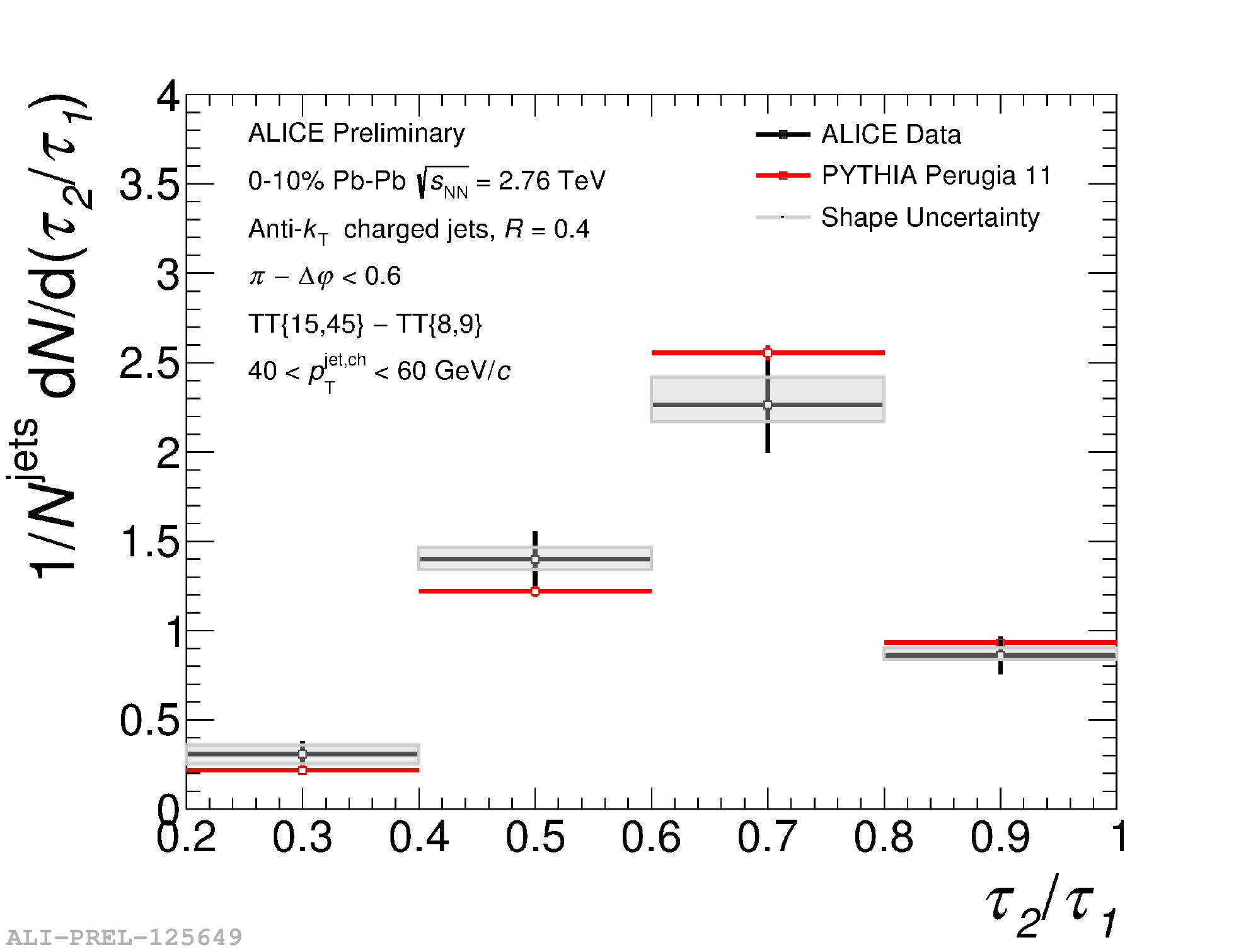}
\caption{Distribution of $\tau_2/\tau_1$ for \antikt\ charged jets with $R=0.4$ in \PbPb\ collisions at $\snn=2.76$~TeV in the $0-10$\% most central events. 
ALICE data points (black) are compared with a PYTHIA6 Perugia-11 simulation (red).}
\label{fig:nsubjettiness}
\end{figure}

Jets are first reconstructed with the \antikt\ algorithm as implemented in the \texttt{FastJet}\cite{Cacciari:2012a} software package.
Tracks with $p_{\rm T}>0.15$~\GeVc\ and $|\eta|<0.9$ identified in the central tracking system
are used for jet reconstruction (\emph{charged jets}). 
Identified jets are reclustered using the exclusive-$k_{\rm T}$ algorithm. In order to obtain the two antenna axes of the jet (first splitting in vaccum) 
the last step in the clustering procedure is unwinded.
The average uncorrelated background is subtracted from the measured nsubjettiness with a constituent-based subtraction~\cite{Berta:2014}.
A PYTHIA6+GEANT3 simulation is used to estimate the detector effects on the jet \pt\ and on $\tau_2/\tau_1$ and a correction is applied using the 2D Bayesian method~\cite{Dagostini:1995}.
The hadron-jet coincidence technique described in~\cite{ALICE:2015g} is used to reject combinatorial jets from the measured sample. 
Instead of the absolute cross section, which suffers from a large combinatorial contamination, the difference in the yields of jets recoiling from two high-\pt\ trigger hadron classes is reported. 
The trigger hadron \pt\ classes are $15-45$~\GeVc\ (signal class) and $8-9$~\GeVc\ (reference class). The distribution of $\tau_2/\tau_1$ is shown in Fig.~\ref{fig:nsubjettiness} for jets with $40<p_{\rm T,ch.jet}<60$~\GeVc.
The measured distribution is compared with a PYTHIA6 Perugia-11 simulation which agrees with the data within the uncertainties. 
Jets that have $\tau_2/\tau_1\approx0$ are the best candidates to look for coherent effects in the jet energy loss; 
in order to get a better sensitivity one should look differentially into the opening angle $\theta$ between the two subjets for this class of two-pronged jets.
Furthermore, more detailed studies are needed to fully assess the role of the background in the determination of the two subjets. 
These developments are being pursued by the ALICE
Collaboration and new results will be made available in the near future.

\subsection{Charm Elliptic Flow}
\begin{figure}[tb]
\centering
\begin{subfigure}[b]{0.495\textwidth}
  \centering
  \includegraphics[width=1.0\linewidth]{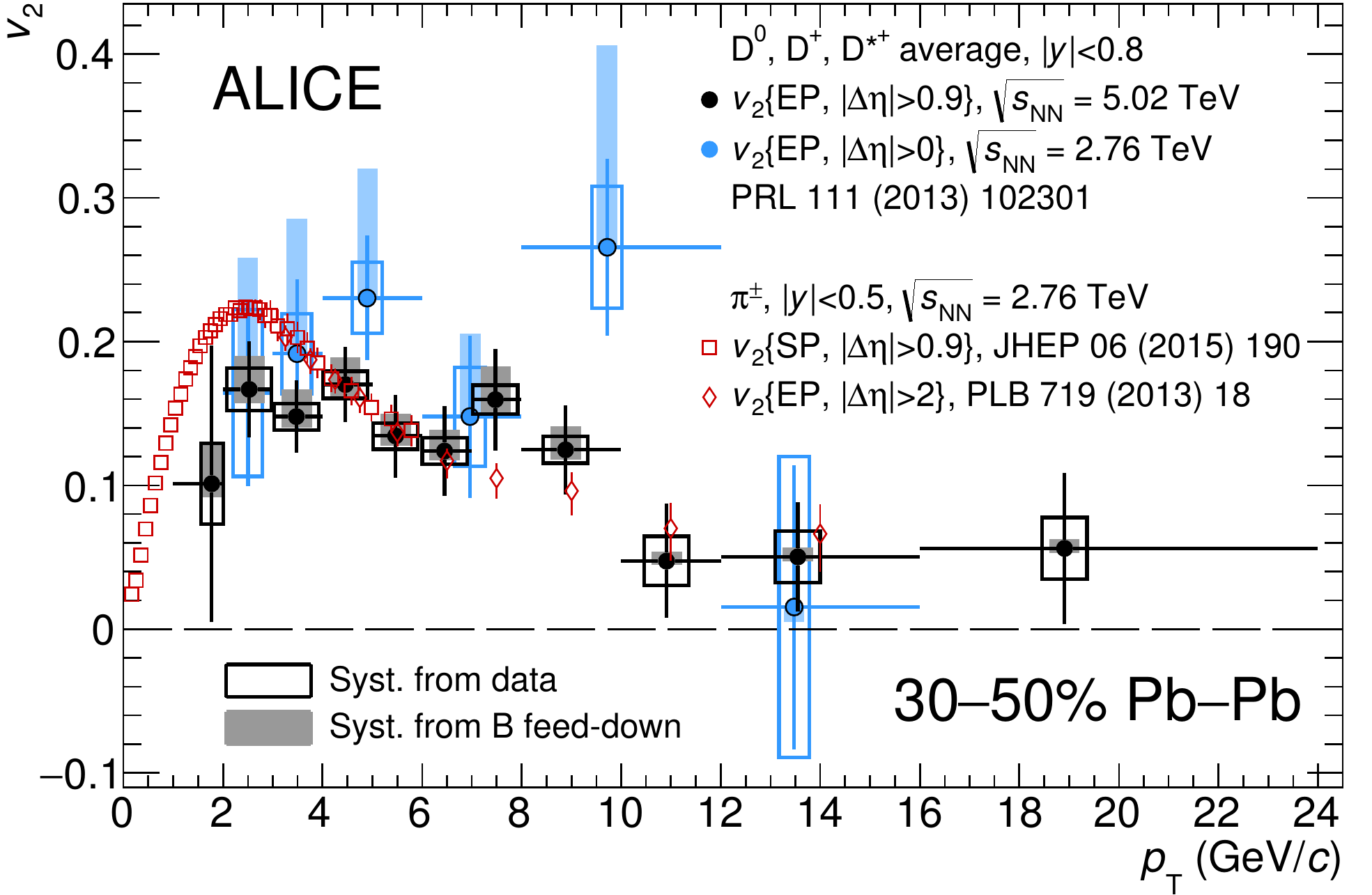}
  \caption{D-meson~\cite{ALICE:2017c}.}
  \label{fig:v2d0}
\end{subfigure} \quad
\begin{subfigure}[b]{0.475\textwidth}
  \centering
  \includegraphics[width=1.0\linewidth]{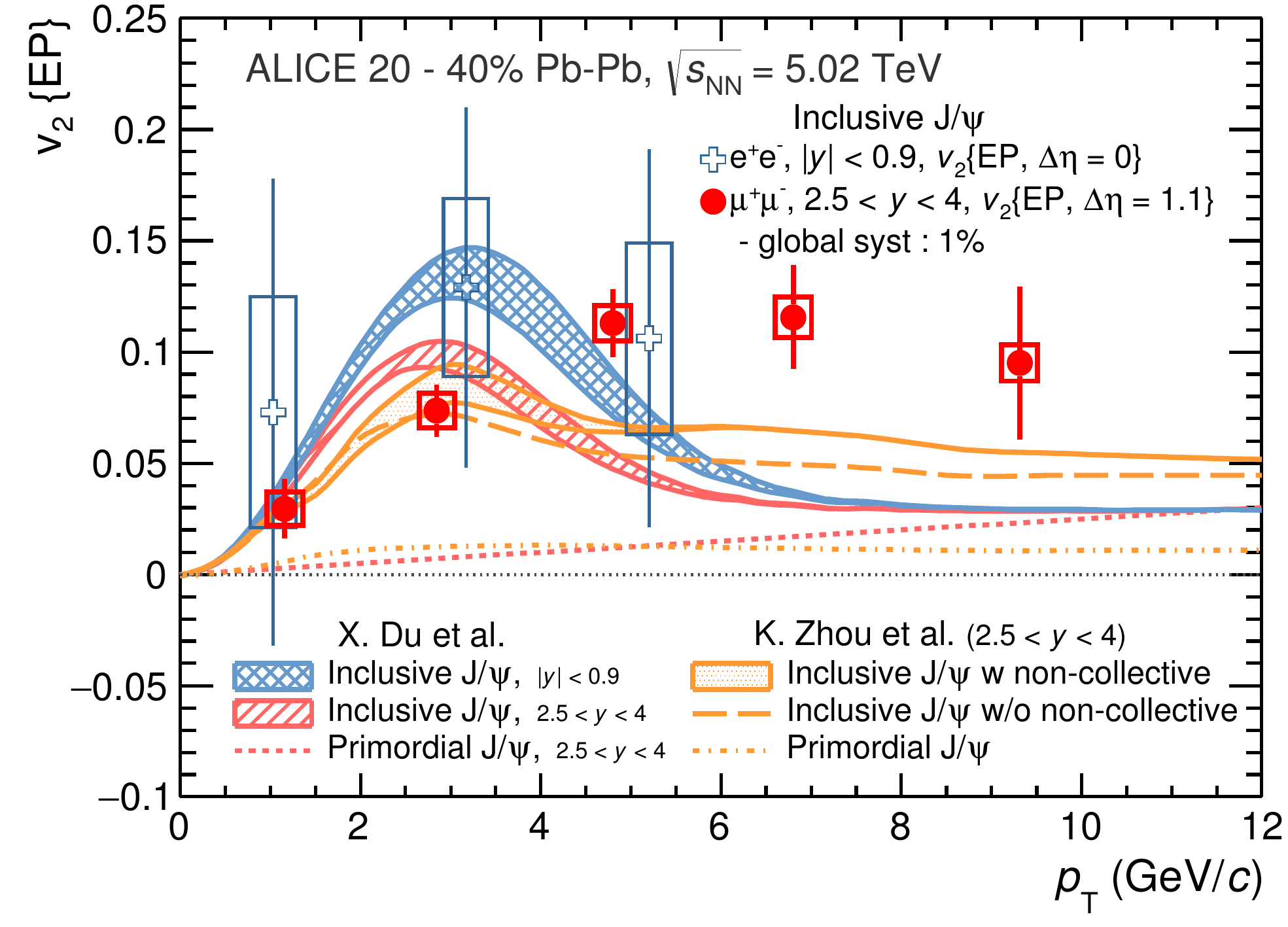}
  \caption{\jpsi~\cite{ALICE:2017d}.}
  \label{fig:v2jpsi}
\end{subfigure} \quad
\caption{Left: elliptic-flow parameter $v_2$ of \Dzero\ mesons (black) in semi-central \PbPb\ collisions at $\snn=5.02$~TeV compared with the same measurement at $\snn=2.76$~TeV (blue)
and with the elliptic flow of pions (red). Right: elliptic-flow parameter $v_2$ of \jpsi\ at mid-rapidity (red) and forward-rapidity (blue) compared with theoretical calculations~\cite{Zhou:2014, Du:2015} in the same collision system.}
\end{figure}

The azimuthal anisotropy of D mesons is characterized by $v_2= \left<\cos 2(\phi - \psi_2)\right>$, where $\phi$ is the D-meson azimuthal angle and $\psi_2$ is the symmetry plane of the second-order
harmonic in the Fourier decomposition of the azimuthal distribution of the particles produced in the event.
This coefficient is usually referred to as \emph{elliptic flow} because its existence is ascribed to the initial asymmetric spatial configuration of the colliding-nuclei overlap region, characteristic of semi-central heavy-ion collisions.
The spatial asymmetry gives rise to azimuthal asymmetric pressure gradients.
The measurement of the elliptic flow of D mesons provides important insights into the interactions of charm quarks with the
medium constituents. At low \pt, the D-meson $v_2$ offers the unique opportunity to test whether also charm quarks participate in
the collective expansion dynamics and possibly thermalize in the medium~\cite{Greco:2003}. At high \pt, it provides constraints on the path-length dependence of parton energy loss~\cite{Gyulassy:2000}.
The measurement of the $v_2$ of D mesons has played an important role in challenging
the theoretical models~\cite{Das:2015}, so it is paramount to measure it with the best precision possible.
Figure~\ref{fig:v2d0} shows the elliptic flow of D mesons in semi-central \PbPb\ collisions at $\snn=5.02$~TeV as a function of \pt.
Non-flow contributions to the $v_2$ coming from decays and jets, are strongly
reduced by the separation of at least 0.9 units in $\eta$ between the D mesons and the particles used
to estimate the event plane. 
The values of $v_2$ are greater than zero and compatible with the corresponding values for pions within uncertainties in the range $2<\pt<24$~\GeVc.

Hidden-charm mesons have an equally important role in the characterization of the QGP.
The observed \jpsi\ yield in heavy-ion collisions has been interpreted as being the result of two competing effects: dissociation of the \ccbar\ pairs in the QGP; and generation in the QGP from pairs of charm and anti-charm quarks coming from initially independent hard scatterings.
The observed non-zero $v_2$ of the \jpsi, shown in Fig.~\ref{fig:v2jpsi}, is an indication of a large fraction of \jpsi\ generated in the QGP out of thermalized pairs of charm-anticharm quarks.

\section{\pPb\ collisions}

\subsection{Hadron-Jet Cross Section} 
A challenge presented by small systems, \pp\ or \pPb, lies in the weak correlation between the EA and the collision geometry,
as various conservation laws and fluctuations play an important role and induce bias in the geometric modelling~\cite{ALICE:2014e}. 
This issue becomes even more prominent when measuring hard processes, since they introduce additional biases on EA estimators.
This bias can be partially avoided by constructing observables which allow to identify medium-induced effects without the need to know the relation between EA and collision geometry. 
An example is the semi-inclusive cross section of the coincidence of high-\pt\ hadrons and recoiling jets.
This cross section is defined as the ratio of two cross sections as in Eq.~\ref{eq:hjet}. Here, $\sigma^{{\rm p+Pb (p+p)}\rightarrow{\rm h+X}}$ is the inclusive single-hadron cross
section, $\mathrm{d}\sigma^{{\rm p+Pb (p+p)}\rightarrow{\rm h+jet+X}}/ \mathrm{d}p_{\rm T,jet}^{\rm ch}$ is the hadron+jet \pt-differential inclusive cross section.
$T_{\rm pPb}$ is the nuclear thickness of the Pb nucleus traversed by the proton, which depends on the collision geometry and cancels in the equation.
\begin{equation}
\frac{1}{\sigma^{{\rm p+Pb}\rightarrow{\rm h+X}}} \frac{\mathrm{d}\sigma^{{\rm p+Pb}\rightarrow{\rm h+jet+X}}}{\mathrm{d}p_{\rm T,jet}^{\rm ch}} = 
\frac{1}{\cancel{T_{\rm pPb}}\sigma^{{\rm p+p}\rightarrow{\rm h+X}}} \frac{\cancel{T_{\rm pPb}}\mathrm{d}\sigma^{{\rm p+p}\rightarrow{\rm h+jet+X}}}{\mathrm{d}p_{\rm T,jet}^{\rm ch}}
\label{eq:hjet}
\end{equation}
Jets are reconstructed using only charged tracks using the \antikt\ algorithm with a resolution parameter $R=0.4$.
As for the nsubjettiness measurement described in Section~\ref{sect:nsubjettiness}, in order to remove the contamination coming from combinatorial jets the difference between the yields
for two trigger hadron classes are reported~\cite{ALICE:2015g}. In this case, the trigger hadron \pt\ classes are $12-50$~\GeVc\ (signal class) and $6-7$~\GeVc\ (reference class).
The raw yields are unfolded to correct for instrumental effects and local background fluctuations.

\begin{figure}[tb]
\centering
\includegraphics[width=.5\textwidth]{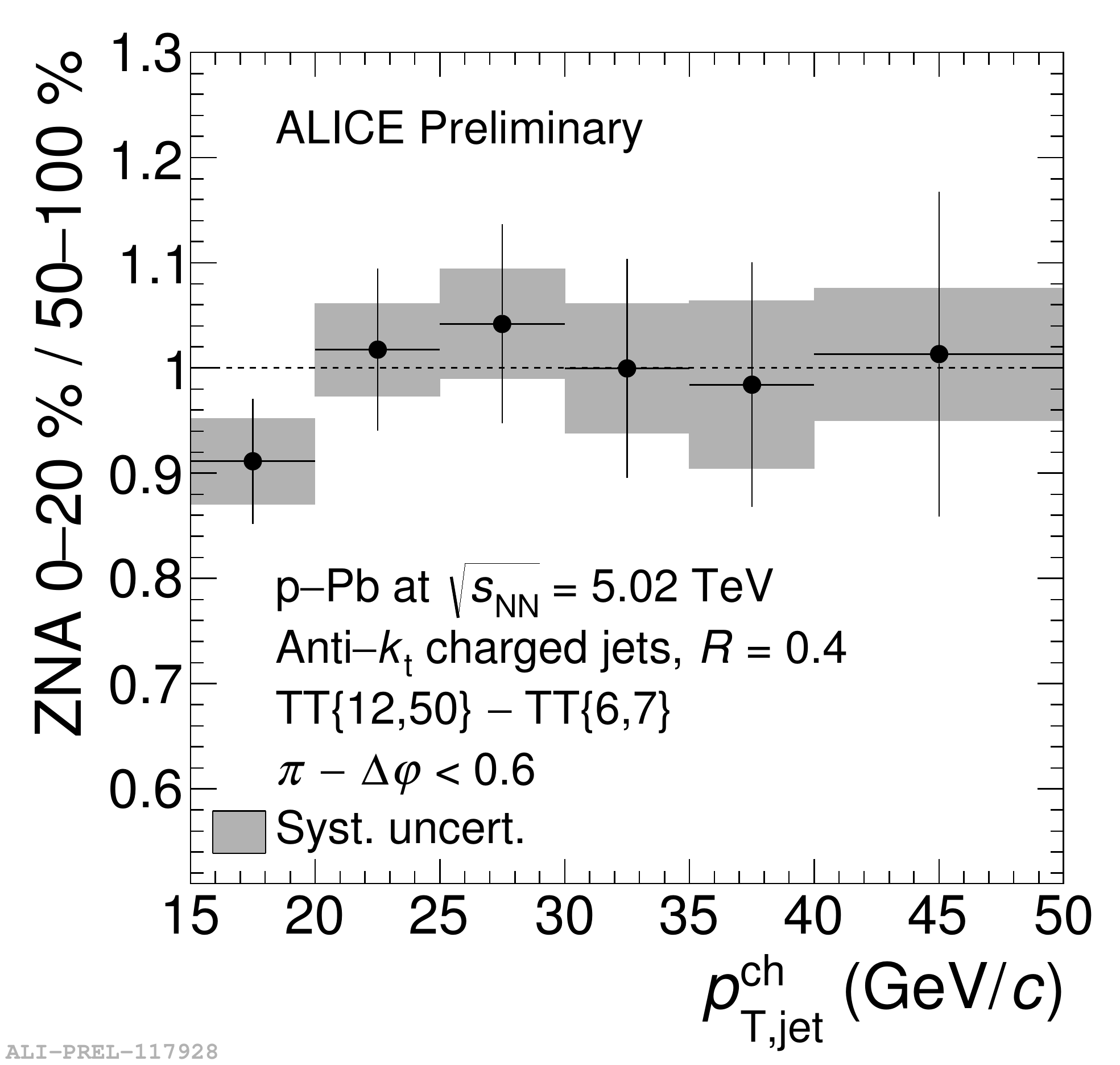}
\caption{Ratio of the semi-inclusive \pt-differential cross sections of hadron+jet in the $0-20$\% largest EA over $50-100$\% \pPb\ collisions at $\snn=5.02$~TeV.}
\label{fig:rppbhjet}
\end{figure}

The \pPb\ events are divided in EA classes based on the energy measured by the Zero-Degree Calorimeter positioned in the forward direction of the Pb-going side.
Figure~\ref{fig:rppbhjet} shows the ratio of the semi-inclusive hadron-jet cross section in the $0-20$\% events with largest EA over the $50-100$\%
in \pPb\ collisions at $\snn=5.02$~TeV.
The ratio is consistent with unity within uncertainty, which puts severe constraints on jet energy loss in high EA \pPb\ collisions.

\subsection{Jet Splitting Function}
\begin{figure}[tb]
\centering
\includegraphics[width=.85\textwidth]{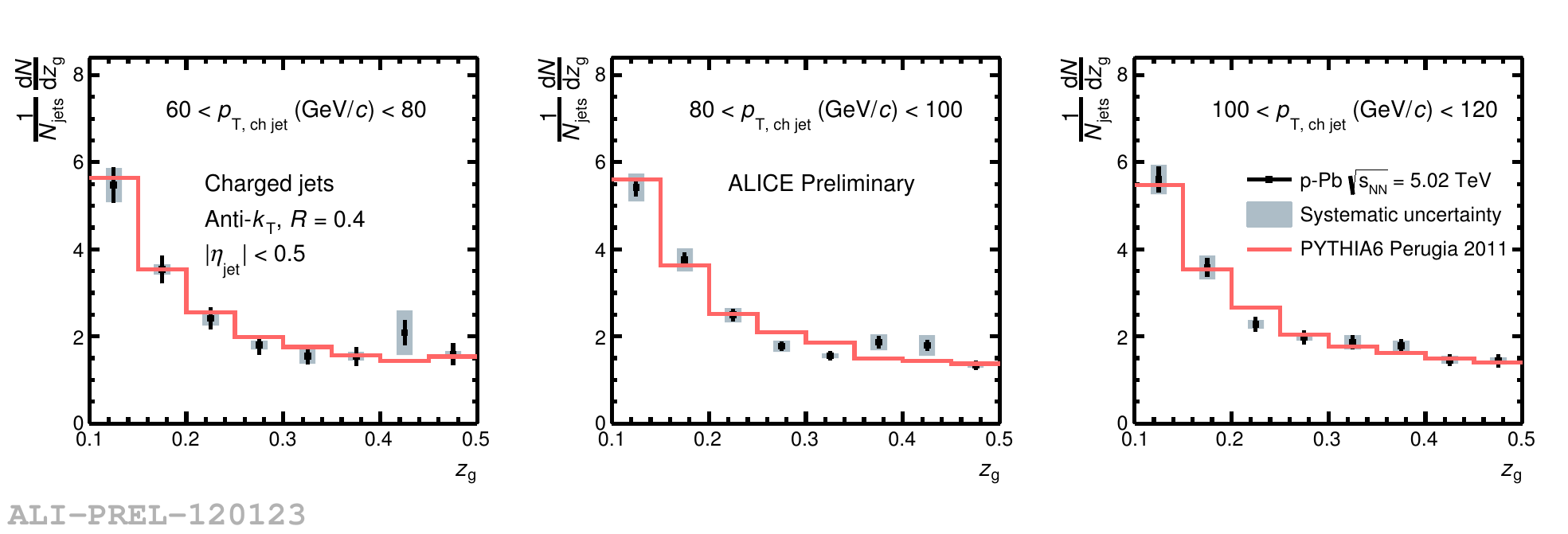}
\caption{Distribution of $z_{\rm g}$ in \pPb\ collisions at $\snn=5.02$~TeV in three jet \pt\ ranges, $60-80$ (left), $80-100$ (center) and $100-120$ (right) \GeVc.
The red curve represents a PYTHIA6 Perugia-11 simulation.}
\label{fig:zgppb}
\end{figure}

The jet splitting function $z_{\rm g}$ is defined in Eq.~\ref{eq:zg}, where $p_{\rm T,1}$ and $p_{\rm T,2}$ are the transverse momenta
of the two hard subjets.
\begin{equation}
z_{\rm g} = \frac{\min({p_{\rm T,1},p_{\rm T,2}})}{p_{\rm T,1}+p_{\rm T,2}}
\label{eq:zg}
\end{equation}
In order to find the two hard subjets a ``soft drop'' algorithm is employed~\cite{Larkoski:2014, Larkoski:2015}. Jets are first identified via the \antikt\ algorithm, then reclustered using Cambridge/Aachen algorithm.
The last step of the clustering procedure is unwinded and the soft drop condition $z_{\rm g} > z_{\rm g, min}$ is applied. If the soft drop condition is not satisfied the softer subjet is dropped.
The procedure continues by unwinding the previous clustering step until the soft drop condition is satisfied or the last clustering step is reached (in which case the jet is rejected).
When applied to jets in heavy-ion collisions, this jet shape observable is sensitive to nuclear effects occurring in the early stages of the parton shower evolution.
CMS has recently reported a modification of the jet splitting function in central \PbPb\ collisions at $\snn=5.02$~TeV compared with
\pp\ collisions at the same energy~\cite{CMS:2017a}.
STAR has performed a similar measurement in Au--Au collisions at $\snn=200$~GeV~\cite{Kauder:2017}.
At this conference, ALICE has reported a measurement of the jet splitting function in \pPb\ collisions at $\snn=5.02$~TeV to investigate possible cold nuclear matter effects.
Figure~\ref{fig:zgppb} shows the distribution of $z_{\rm g}$ of charged jets in \pPb\ collisions at $\snn=5.02$~TeV in three jet \pt\ ranges, $60-80$, $80-100$ and $100-120$~\GeVc. 
The distributions are corrected for instrumental effects as well as local background fluctuations with a Bayesian 2D unfolding procedure.
No modification is observed when comparing the data with a PYTHIA6 Perugia-11 simulation.

\section{Conclusions}
The picture that emerges from the most recent measurements from ALICE using Run-2 data confirms earlier results from Run-1 with significantly better precision.
Jets and heavy-flavor mesons are strongly suppressed in central \PbPb\ collisions compared to \pp\ collisions. Both open and hidden charm mesons show a non-zero $v_2$ which indicates 
that charm quarks participate in the collective dynamic expansion with the medium.
ALICE has not observed any modification in the internal substructure of jets so far. Work is in progress to extend the $z_{\rm g}$ measurement to \PbPb\ collisions
in order to provide an independent confirmation of the modification reported by the CMS Collaboration and possibly extend the measurement to a different kinematic range.
In order to investigate the details of the energy loss of charm quarks in the QGP, ALICE is working on a measurement of charm jets tagged using D mesons.
A preliminary cross section of \Dzero-tagged jets in \pp\ collisions have been recently reported~\cite{Aiola:2017} and the measurement is being extended to both \pPb\ and \PbPb\ collision systems.

In \pPb\ collisions ALICE has confirmed that there is no evidence of jet quenching due to either cold or hot nuclear matter effects.
In particular, the self-normalized ratio of the semi-inclusive cross section of hadron+jet in high EA over low EA collisions is compatible with unity within uncertainties.
A first look into the internal hard substructure of jets in \pPb\ collisions has also shown that they are compatible with expectations from PYTHIA6.
From the one hand these findings confirm the interpretation of the jet quenching observed in \PbPb\ collisions as the result of the interactions of hard scattered partons
with the QGP; from the other hand they suggest that no QGP is formed in high EA \pPb\ collisions or that the drop of QGP formed in such collisions is too small or dilute to give rise to observable jet quenching effects.
 



\bibliography{biblio}{}
\bibliographystyle{utphys}
 
\end{document}